\documentstyle[11pt,epsfig]{article}
\begin{document}
\def\sqr#1#2{{\vcenter{\vbox{\hrule height.#2pt
        \hbox{\vrule width.#2pt height#1pt \kern#1pt
          \vrule width.#2pt}
        \hrule height.#2pt}}}}
\def\square{\mathchoice\sqr34\sqr34\sqr{2.1}3\sqr{1.5}3}
\def\Square{\mathchoice\sqr67\sqr67\sqr{5.1}3\sqr{1.5}3}
\title{\bf Anisotropic Cosmological Model in Modified Brans--Dicke Theory}
\author{S.M. M. Rasouli\thanks{email: m-rasouli@sbu.ac.ir},\ Mehrdad Farhoudi\thanks{email: m-farhoudi@sbu.ac.ir}\ \ and
        Hamid R. Sepangi\thanks{email: hr-sepangi@sbu.ac.ir} \\{\small Department of Physics, Shahid Beheshti
        University, G. C.,
        Evin, Tehran 19839, Iran}}
\date{\small February 23, 2011}
\maketitle
\begin{abstract}
Recently, it has been shown that a four--dimensional ($4D$)
Brans--Dicke (BD) theory with an effective matter field and a self
interacting potential can be achieved from the vacuum $5D$ BD
field equations, where we refer to as a modified Brans--Dicke
theory (MBDT). We investigate a generalized Bianchi type~I
anisotropic cosmology in $5D$ BD theory, and by employing the
obtained formalism, we derive the induced--matter on any $4D$
hypersurface in the context of the MBDT. We illustrate that if the
usual spatial scale factors are functions of the time while the
scale factor of extra dimension is constant, and the scalar field
depends on the time and the fifth coordinate, then in general, one
will encounter inconsistencies in the field equations. Then, we
assume the scale factors and the scalar field depend on the time
and the extra coordinate as separated variables in the power--law
forms. Hence, we find a few classes of solutions in $5D$
space--time through which, we probe the one which leads to a
generalized Kasner relations among the Kasner parameters. The
induced scalar potential is found to be in the power--law or in
the logarithmic form, however, for a constant scaler field and
even when the scalar field only depends on the fifth coordinate,
it vanishes. The conservation law is indeed valid in this MBDT
approach for the derived induced energy--momentum tensor (EMT). We
proceed our investigations for a few cosmological quantities,
where for simplicity we assume the metric and the scalar field are
functions of the time. Hence, the EMT satisfies the barotropic
equation of state, and the model indicates that constant mean
Hubble parameter is~not allowed. Thus, by appealing to the
variation of Hubble parameter, we assume a fixed deceleration
parameter, and set the evolution of the quantities with respect to
the fixed deceleration, the BD coupling and the state parameters.
The WEC allows a shrinking extra dimension for a decelerating
expanding universe that, in the constant scalar field, evolves the
same as the flat FRW space--time in GR. The quantities for the
stiff fluid and the radiation dominated universe indicate an
expanding universe commenced with a big bang. There is a horizon
for each of the fluids, and the rate of expansion slows down by
the time. The allowed ranges of the deceleration and the BD
coupling parameters have been obtained, and the model gives an
empty universe when the time goes to infinity.
\end{abstract}
\bigskip
\section{Introduction}
\indent

The original version of the Kaluza--Klein
theory~\cite{Kaluza21,Klein26} assumes, as a postulate, that the
fifth dimension is compact. However, a particular variety of the
Kaluza--Klein theory which is non--compact has also been proposed
by Wesson~\cite{ponce-wesson-92}--~\cite{5Dwesson06}, known as the
induced--matter theory~({\bf IMT}) and also as the
space--time--matter theory. This theory is based on the
Campbell-–Magaard
theorem~\cite{Lidsey-Romero-Tavakol-Rippl1997}--~\cite{Seahra-Wesson2003},
which states that any analytic $D$--dimensional Riemannian
manifold can be locally embedded in a $(D +1)$--dimensional
Ricci--flat space--time. The core of IMT applied to general
relativity yields the $4D$ Einstein equation as a subset of a $5D$
Einstein vacuum equation with an effective or induced EMT which
contains the classical properties of matter. The route usually
taken in the IMT is to adapt appropriate coordinates that yield
the induced--matter consisted with the observations. A generalized
approach of the IMT, which relates a vacuum $(D+1)$--dimensional
space--time to a $D$--dimensional theory with source, has also
been proposed~\cite{rippl-romero-tavakol-95}.

Many exact solutions have been found for $5D$ Einstein equations
in the IMT by allowing the metric to depend on an extra coordinate
and by regarding effects of this extra dimension as an induced--
matter in a $4D$ space--time, see, e.g., Ref.~\cite{WPLM96}. Some
of these solutions have remarkable mathematical and physical
properties. For example, the solutions which are flat in $5D$
space--time but curved in four dimensions provide suitable
descriptions of, ordinary matter in the present and the early
universe~\cite{W94,DRJ09}, and of waves in the inflationary
phase~\cite{BW96}. The new solutions in $5D$ space--time have
significant applications and relatively more complicated geometry
than $5D$ Riemannian one which may be needed to explain the gauge
properties of elementary particles.

Recently, another generalization of the IMT has been proposed for
the BD theory as a fundamental underlying
theory~\cite{aguilar-romero-barros-2007}--~\cite{Ponce2-10},
though the effective equations and quantities in $4D$
hypersurfaces obtained in Ref.~\cite{aguilar-romero-barros-2007}
is different from those in Refs.~\cite{Ponce1-10,Ponce2-10}.
Further application of the former approach has been performed
in~\cite{BFS10}, though in this work we employ the latter approach
and also pursue a $D$--dimensional version of it in another
work~\cite{RF}. In Ref.~\cite{Ponce2-10}, the MBDT equations are
obtained from $5D$ BD equations in which the induced--matter can
still be viewed as being generated by a fifth dimension. The
effective induced--matter\rlap,\footnote{In this work, we assume
that there is no matter in the $5D$ space--time.}\
 $T^{({\rm EMT})}_{\mu\nu}$, is composed of two parts. One part,
which we present by $T^{(\varphi)}_{\mu\nu}$, depends on the BD
scalar field, $\varphi$, and its derivatives with respect to the
fifth coordinate. Another part, which we denote by $T^{({\rm
IMT})}_{\mu\nu}$, is induced from the extra dimensional part of
the metric and corresponds to the related one in the IMT. An
induced scalar potential is also produced geometrically from the
$5D$ space--time. As the BD theory can be considered as a modified
version of general relativity, one may also regard the MBDT as a
modified version of the IMT.

We present and discuss the exact solutions of a generalized
Bianchi type~I model in the MBDT. The Bianchi models are
cosmological models that have spatial homogeneous sections
invariant under the action of a three--dimensional Lie group, but
not necessarily isotropic ones. The Bianchi type~I model
corresponds to the flat spatial sections as a homogeneous
universe. It generalizes the flat Friedmann--Robertson--Walker
model (the simplest spatial homogeneous one), which allows for
anisotropy.

The Bianchi type~I model for the vacuum solutions are called the
Kasner solutions due to E. Kasner who first found such solutions
in 1921~\cite{kasner-21}. When the EMT vanishes, the Einstein
tensor, due to the Einstein equation, has to vanish as well, but
it is~not zero in the BD theory. The BD theory for the various
Bianchi types~I--IX models have been considered in the literature,
see, e.g., Refs.~\cite{RS83}--~\cite{J.L -M.Nahmad-01}. Also, the
Bianchi type~I models with non--vanishing cosmological constant as
a function of the scalar field, $\Lambda(\varphi)$, have been
investigated in the framework of BD theory~\cite{b-s-82}. In
addition, the Kasner cosmologies have been considered in the BD
theory in four
dimensions~\cite{johri-83,banerjee-banerjee-santos-85}. Various
vacuum solutions as well as some dust solutions of the Bianchi
type~I models with cosmological constant have been found in the
standard BD theory~\cite{lorenz-84}. Furthermore, the Bianchi
type~I models have been studied in the context of
IMT~\cite{H01,PW08} and in the quantum cosmology~\cite{C07,VS07}.

In the next section, we give a brief review of the MBDT. In
Section~$3$, we investigate a generalized Bianchi type~I model in
$5D$ space-time, derive the corresponding field equations and
meanwhile, point to an inconsistency that appears due to the
assumption made in the literature. In the rest of this section, we
solve the equations in a general form under the assumption of
separation of variables, and obtain five classes of solutions in
five dimensions. Among these five classes, we are interested in a
case (the Case II) which is a generalized Kasner model in $5D$
space-times. Then, solving the corresponding equations leads to a
few relations among the generalized Kasner parameters. In
Section~$4$, by applying the effective equations achieved in
Section~$2$, we derive the scalar potential, the energy density
and pressure of the induced--matter and also discuss the
properties of some cosmological quantities. Finally, we draw
conclusions in Section~$5$.

\section{Modified BD Theory From $5D$ Vacuum One}
\indent

In the present section, we launch a brief review on the MBDT, the
core idea of
Refs.~\cite{aguilar-romero-barros-2007}--~\cite{Ponce2-10} while
focusing on the latter ones approach. The action of $4D$ BD
theory~\cite{brans-dicke-61} with a self interacting
potential\footnote{In the literature, in order to obtain
accelerating universes, inclusion of such potentials has been
considered in {\it priori} by hand.}\
 and a matter field is
\begin{equation}\label{4d-action}
S^{(4)}[g_{\alpha\beta}, \varphi]=\int d^{4}x\sqrt{-g}\Bigg\{\frac{1}{16\pi}\left[\varphi
R^{(4)}-\frac{\omega}{\varphi}g^{\alpha\beta}(D_\alpha\varphi)(D_\beta\varphi)-V(\varphi)\right]
+L_{\rm matt}^{(4)}\Bigg\},
\end{equation}
where\footnote{In order to have a non--ghost scalar field in the
conformally related Einstein frame, i.e. a field with a positive
kinetic energy term in that frame, the BD coupling parameter must
be restricted to $\omega>-(D-1)/(D-2)$, where $D$ is the dimension
of space--time~\cite{Freund1982}--~\cite{bkm04}.}\
 $\omega>-3/2$,\, $c=1$, the $D_\alpha$ is the covariant derivative in $4D$
space--time and the Greek indices run from zero to three. The  BD
scalar field plays a role analogous to the inverse of the Newton
gravitational constant, the $\omega$ is a dimensionless coupling
parameter, the Lagrangian $ L^{(4)}_{\rm matt}$ depends on the
metric and other matter fields except $\varphi$, and we have
chosen $c=1$. The variation of action (\ref{4d-action}) with
respect to the metric and the scalar field gives the field
equations
\begin{equation}\label{4d-field-eq1}
G^{(4)}_{\alpha\beta}=\frac{8\pi}{\varphi}T^{(4)}_{\alpha\beta}
+\frac{\omega}{\varphi^{2}}
\left[(D_\alpha\varphi)(D_\beta\varphi)
-\frac{1}{2}g_{\alpha\beta}(D_\sigma\varphi)(D^\sigma\varphi)\right]+\frac{1}{\varphi}
\left[D_{\alpha}D_{\beta}\varphi-g_{\alpha\beta} \left(D^2\varphi
+\frac{1}{2}V(\varphi)\right)\right]
\end{equation}
and
\begin{equation}\label{4d-field-eq2}
D^2\varphi=\frac{8\pi
}{2\omega+3}T^{(4)}+\frac{1}{2\omega+3}\left[\varphi\frac{dV(\varphi)}{d\varphi}-2V(\varphi)\right]\,,
\end{equation}
where $\delta\left(L_{\rm matt}^{(4)}\sqrt{-g}\,\right) \equiv
-{1\over 2}\sqrt{-g}\ T^{(4)}_{\alpha\beta}\, \delta
g^{\alpha\beta}$ and $D^2\equiv D^\alpha D_\alpha$. In the
standard $4D$ BD theory, one is led to the accelerating cosmos in
a number of ways. For example, assuming the coupling parameter
$\omega$ as a function of the time~\cite{banerjee-pavon-01}, or
introducing a time--dependent cosmological term~\cite{MC007} and
or adding a scalar potential to the Lagrangian by means of an {\it
ad hoc} assumption~\cite{sen-sen-01}. However, by employing the
Wesson idea for the BD theory, it has been
shown~\cite{Ponce1-10,Ponce2-10} that such a scalar potential can
geometrically be induced due to the extra fifth dimension in the
MBDT. In this approach, the scalar potential has the desired
properties and leads to an accelerating universe. In what follows
we review the approach of Ref.~\cite{Ponce2-10} and then, we
employ this formalism in the Bianchi type~I anisotropic model.

The action for a BD theory in $5D$ space--time, without a self
interacting potential, can analogously be written as~\cite{QMHY05}
\begin{equation}\label{5d-action}
 S^{(5)}[g_{_{AB}}, \varphi]=\int\,d^{5}x
 \sqrt{\Bigl|g^{(5)}\Bigr|}\Bigg\{\frac{1}{16\pi}\left[\varphi R^{(5)}
-\frac{\omega}{\varphi}
g^{AB}(\nabla_{A}\varphi)(\nabla_{B}\varphi)\right]+L^{(5)}_{\rm
matt}\Bigg\},
\end{equation}
where $\omega>-4/3$, the Latin indices run from zero to four, the
$\nabla_A$ is the covariant derivative in $5D$ space--time,
$R^{(5)}$, $g^{(5)}$ and $L^{(5)}_{\rm matt}$ are analogs of the
$4D$ quantities. The field equations from action~(\ref{5d-action})
obviously are
\begin{equation}\label{5d-field-eq1}
G^{(5)}_{AB}=\frac{8\pi}{\varphi}T^{(5)}_{AB}+\frac{\omega}{\varphi^{2}}
\left[(\nabla_{A}\varphi)(\nabla_{B}\varphi)-\frac{1}{2}g_{AB}(\nabla^{C}\varphi)(\nabla_{C}\varphi)\right]
+\frac{1}{\varphi}\left(\nabla_A\nabla_B\varphi-g_{AB}\nabla^2\varphi\right)
\end{equation}
and
\begin{equation}\label{5d-field-eq2}
\nabla^2\varphi=\frac{8\pi}{3\omega+4}\,T^{(5)}\,,
\end{equation}
where $\nabla^2\equiv\nabla_A\nabla^A$. The vacuum case is defined
as a configuration where there is no matter, except the scalar
field, in $5D$ space--time. Thus, by setting $T^{(5)}_{AB}=0$ and
consequently $T^{(5)}\equiv g^{AB}T^{(5)}_{AB}=0$,
Eqs.~(\ref{5d-field-eq1}) and~(\ref{5d-field-eq2}) read
\begin{equation}\label{5d-vacuum1}
G^{(5)}_{AB}=\frac{\omega}{\varphi^{2}}
\left[(\nabla_{A}\varphi)(\nabla_{B}\varphi)-\frac{1}{2}g_{AB}(\nabla^{C}\varphi)(\nabla_{C}\varphi)\right]
+\frac{1}{\varphi}\left(\nabla_A\nabla_B\varphi-g_{AB}\nabla^2\varphi\right)
\end{equation}
and
\begin{equation}\label{5d-vacuum2}
\nabla^2\varphi=0.
\end{equation}

Now, following the idea of the IMT approach, one can extract the
effective $4D$ equations. Suppose that the $5D$ line element can be
written in local coordinates $x^{A}=(x^{\mu}, l)$ as
\begin{equation}\label{general-metric}
dS^{2}=g_{AB}(x^{C})dx^{A}dx^{B}=g_{\mu\nu}(x^{C})dx^{\mu}dx^{\nu}
+g_{44}(x^{C})dl^{2}\,,
\end{equation}
where the $l$ represents the fifth coordinate. This metric is
restrictive, but one limits oneself to it for reasons of
simplicity. It is also favorable to write the fifth component of
the metric as $g_{44}(x^{C})=\epsilon h^2(x^C)$, where the
$h(x^C)$ is a scalar field and $\epsilon=\pm 1$ indicates the sign
of the fifth dimension. Assuming the $5D$ space--time is foliated
by a family of hypersurfaces $\Sigma$ defined by fixed values of
the fifth coordinate, where the induced metric on each
hypersurface, e.g. $\Sigma_0(l=l_0)$, is given by
\begin{equation}\label{induced-metric}
ds^{2}=g^{(5)}_{\mu\nu}(x^{\alpha}, l_{o})dx^{\mu}dx^{\nu}\equiv
g_{\mu\nu}dx^{\mu}dx^{\nu}.
\end{equation}
The $4D$ orthogonal unit vector along the extra dimension is
written as
\begin{equation}\label{extradimension}
n^A=\frac{\delta^{A}_{4}}{h} \qquad {\rm where} \qquad
n_An^A=\epsilon.
\end{equation}

Dimensional reduction of Eqs.~(\ref{5d-vacuum1}) and
(\ref{5d-vacuum2}) yields the effective field equations in $4D$
hypersurfaces. Actually, reduction of the fundamental field
Eq.~(\ref{5d-vacuum1}) and the second field
equation~(\ref{5d-vacuum2}) on the hypersurface $\Sigma_0$ can be
written~\cite{Ponce2-10} exactly the same as the usual $4D$ BD
field Eqs.~(\ref{4d-field-eq1}) and~(\ref{4d-field-eq2}). However
now, the term $T_{\alpha\beta}^{(4)}$, that for clarity we
indicate it by $T_{\alpha\beta}^{({\rm EMT})}$, is the induced EMT
of the effective $4D$ modified BD theory and has the form
\begin{equation}\label{bd-energy-momentum}
T_{\alpha\beta}^{^{(\rm EMT)}}\equiv T_{\alpha\beta}^{^{(\rm
\varphi)}}+ T_{\alpha\beta}^{^{\rm (IMT)}}\,,
\end{equation}
where
\begin{equation}\label{ph-energy-momentum}
\frac{8\pi}{\varphi}T_{\alpha\beta}^{(\varphi)}\equiv
\frac{\epsilon \varphi'}{2\varphi h^2}\left[g'_{\alpha\beta}
+g_{\alpha\beta}\left(\frac{\omega\varphi'}{\varphi}
-g^{\mu\nu}g'_{\mu\nu}\right)\right]+\frac{1}{2\varphi}g_{\alpha\beta}V(\varphi)
\end{equation}
and
\begin{equation}\label{induced-energy-momentum}
\frac{8\pi}{\varphi}T_{\alpha\beta}^{^{(\rm IMT)}}\equiv
\frac{D_\alpha D_\beta h}{h}-
\frac{\epsilon}{2h^2}\left\{\frac{h'g'_{\alpha\beta}}{h}
-g''_{\alpha\beta}+g^{\mu\nu}g'_{\alpha\mu}g'_{\beta\nu}
-\frac{g^{\mu\nu}g'_{\mu\nu}g'_{\alpha\beta}}{2}+
\frac{g_{\alpha\beta}}{4}\left[g'^{\mu\nu}g'_{\mu\nu}+(g^{\mu\nu}g'_{\mu\nu})^2\right]\right\}.
\end{equation}
The prime denotes partial derivative with respect to the extra
coordinate $l$. The first part of the induced EMT comes from the
scalar field and its derivatives with respect to the fifth
coordinate. The second part of $T_{\alpha\beta}^{^{\rm (EMT)}}$,
which is also induced geometrically from the extra dimensional
part of the metric, is a modified version of the corresponding one
in the IMT approach of Ref.~\cite{ponce-wesson-92}.

Furthermore, in this formalism, these induced
Eqs.~(\ref{4d-field-eq1}) and~(\ref{4d-field-eq2}) can also be
derived from an action similar to action (\ref{4d-action}). Hence,
the induced EMT satisfies the conservation law $D_\alpha
T^{\alpha(\rm EMT)}_{\,\,\,\beta}=0$ as in the standard BD theory
in the Jordan frame. Also, the $V(\varphi)$ is an induced scalar
potential which is defined (or fixed) by the procedure rather than
to be introduced {\it priori} by hand. That is, on the
hypersurface $\Sigma_0$ one has~\cite{Ponce2-10}
\begin{eqnarray}\label{gen.potential.eq.}
\varphi\frac{dV(\varphi)}{d\varphi}{\Biggr
|}_{_{\Sigma_{o}}}\!\!\!\!\!&\!\!\!\equiv&\!\!\!-2(\omega+1)\left[\frac{h_{,\mu}\varphi^{,\mu}}{h}
+\frac{\epsilon}{h^2}\left(\varphi''-\frac{\varphi'h'}{h}\right)\right]\cr
&&\!\!\!\!-\frac{\epsilon\omega\varphi'}{h^2}\left(\frac{\varphi'}{\varphi}+g^{\mu\nu}g'_{\mu\nu}\right)
+\frac{\epsilon\varphi}{4h^2}\left[g'^{\mu\nu}g'_{\mu\nu}+(g^{\mu\nu}g'_{\mu\nu})^2\right].
\end{eqnarray}

Obviously, in the particular case $\omega=-1$ and $l$ as a cyclic
coordinate, the induced potential, without loss of generality,
vanishes. Thus, to investigate a general version of $4D$ BD theory
with a cyclic extra coordinate, one should assume $\omega\neq-1$.
However, when the metric depends on the extra coordinate, one has,
in general, a non--vanishing scalar potential.

In the following, we delineate a generalized Bianchi type~I model,
as an application of the above formalism in the cosmological
context.

\section{Generalized Bianchi Type~I Metric in $5D$ BD Theory}
\indent

We apply a generalization of the Bianchi type~I anisotropic model
for a $5D$ space--time (with spacelike extra dimension) in a
comoving reference frame as
\begin{equation}\label{bulk-metric}
dS^{2}=-dt^{2}+a^{2}(t, l)dx^{2}+b^{2}(t, l)dy^{2}+c^{2}(t,
l)dz^{2}+h^{2}(t, l)dl^{2}\,,
\end{equation}
where the $t$ is the cosmic time, $(x, y, z)$ are the Cartesian
coordinates, and $a(t, l)$, $b(t, l)$, $c(t, l)$ and $h(t, l)$ are
different cosmological scale factors in each of the four
directions. When one assumes that there is no matter in $5D$
space--time and $\varphi=\varphi(t, l)$, the field equations
(\ref{5d-vacuum1}) and (\ref{5d-vacuum2}), with a little
manipulation, give
\begin{eqnarray}\label{bulk-eq1}
\frac{1}{h^2}\left(-\frac{a'}{a}
\frac{b'}{b}-\frac{a'}{a}\frac{c'}
{c}-\frac{b'}{b}\frac{c'}{c}
+\frac{a'}{a}\frac{h'}{h}+\frac{b'}
{b}\frac{h'}{h}+\frac{c'}{c}\frac{h'}{h}-
\frac{a''}{a}-\frac{b''}
{b}-\frac{c''}{c}\right)
+H_{1}H_{2}+H_{1}H_{3}+H_{1}H_{4}\cr
+H_{2}H_{3}+H_{2}H_{4}+H_{3}H_{4}
=\frac{\omega}{2}\left(\frac{\dot{\varphi}} {\varphi}\right)^{2}
+\frac{\omega}{2h^{2}}\left(\frac{\varphi'}{\varphi}\right)^{2}+\frac{\ddot{\varphi}}{\varphi}\,,
\end{eqnarray}
\begin{eqnarray}\label{bulk-eq2}
\frac{1}{h^{2}}\left(-\frac{b'}{b}\frac{c'}{c}+\frac{b'}{b}\frac{h'}{h}+\frac{c'}{c}\frac{h'}{h}
-\frac{b''}{b}-\frac{c''}{c}\right)+\frac{\ddot{b}}{b}+\frac{\ddot{c}}{c}+\frac{\ddot{h}}{h}+
H_{2}H_{3}+H_{2}H_{4}+H_{3}H_{4}\cr
=-\frac{\omega}{2}\left(\frac{\dot{\varphi}}{\varphi}\right)^{2}+
\frac{\omega}{2h^{2}}\left(\frac{\varphi'}{\varphi}\right)^{2}+H_1\frac{\dot{\varphi
}}{\varphi}-\frac{1}{h^{2}}\frac{a'}{a}\frac{\varphi'}{\varphi}\,,
\end{eqnarray}
\begin{eqnarray}\label{bulk-eq3}
\frac{1}{h^{2}}\left(-\frac{a'}{a}\frac{c'}{c}+\frac{a'}{a}\frac{h'}{h}+\frac{c'}{c}\frac{h'}{h}-\frac{a''}{a}
-\frac{c''}{c}\right)+
\frac{\ddot{a}}{a}+\frac{\ddot{c}}{c}+\frac{\ddot{h}}{h}+H_{1}H_{3}+H_{1}H_{4}+H_{3}H_{4}\cr
=-\frac{\omega}{2}\left(\frac{\dot{\varphi}}{\varphi}\right)^{2}+
\frac{\omega}{2h^{2}}\left(\frac{\varphi'}{\varphi}\right)^{2}+
H_2\frac{\dot{\varphi}}{\varphi}
-\frac{1}{h^{2}}\frac{b'}{b}\frac{\varphi'}{\varphi}\,,
\end{eqnarray}
\begin{eqnarray}\label{bulk-eq4}
\frac{1}{h^{2}}\left(-\frac{a'}{a}\frac{b'}{b}+\frac{a'}{a}\frac{h'}{h}+\frac{b'}{b}\frac{h'}{h}
-\frac{a''}{a}-\frac{b''}{b}\right)
+\frac{\ddot{a}}{a}+\frac{\ddot{b}}{b}+\frac{\ddot{h}}{h}+H_{1}H_{2}+H_{1}H_{4}+H_{2}H_{4}\cr
=-\frac{\omega}{2}\left(\frac{\dot{\varphi}}{\varphi}\right)^{2}+
\frac{\omega}{2h^{2}}\left(\frac{\varphi'}{\varphi}\right)^{2}+
H_3\frac{\dot{\varphi}}{\varphi}
-\frac{1}{h^{2}}\frac{c'}{c}\frac{\varphi'}{\varphi}\,,
\end{eqnarray}
\begin{eqnarray}\label{bulk-eq5}
-\frac{1}{h^{2}}\left(\frac{a'}{a}\frac{b'}{b}+\frac{a'}{a}
\frac{c'}{c}+\frac{b'}{b}\frac{c'}{c}\right)+
\frac{\ddot{a}}{a}+\frac{\ddot{b}}{b}+\frac{\ddot{c}}{c}+H_{1}H_{2}+H_{1}H_{3}+H_{2}H_{3}\cr
=-\frac{\omega}{2}\left(\frac{\dot{\varphi}}{\varphi}\right)^{2}-
\frac{\omega}{2h^{2}}\left(\frac{\varphi'}{\varphi}\right)^{2}+H_{4}\frac{\dot{\varphi}}{\varphi}-
\frac{1}{h^{2}}\frac{\varphi''}{\varphi}
+\frac{h'}{h^{3}}\frac{\varphi'}{\varphi}\,,
\end{eqnarray}
\begin{equation}\label{bulk-eq6}
\frac{(\dot{a})'}{a}+\frac{(\dot{b})'}{b}+\frac{(\dot{c})'}{c}
-H_{4}\left(\frac{a'}{a}+\frac{b'}{b}+\frac{c'}{c}\right)
=-\omega\frac{\dot{\varphi}\varphi'}{\varphi^{2}}
-\frac{(\dot{\varphi})'}{\varphi}+H_{4}\frac{\varphi'}{\varphi}
\end{equation}
and
\begin{equation}\label{bulk-eq7}
-\frac{1}{h^{2}}\left(\frac{a'}{a}+\frac{b'}{b}+\frac{c'}{c}-\frac{h'}{h}\right)\varphi'
+\left(\ddot{\varphi}-\frac{1}{h^{2}}\varphi''\right)
+\Big(H_{1}+H_{2}+H_{3}+H_{4}\Big)\dot{\varphi}=0\,,
\end{equation}
where the dot denotes derivative with respect to the cosmic time,
$H_{1}$, $H_{2}$, $H_{3}$ and $H_{4}$ are the corresponding Hubble
parameters in each of the four directions, respectively.

Meanwhile let us revisit a specific case that has been employed in
the literature. When $h(t,l)$ is a constant and the other scale
factors are functions of the cosmic time only, then the Einstein
tensor, $G_{_{AB}}$, will be a function of time as well. Then, if
one insists that the scalar field, $\varphi$, to remain a function
of the fifth coordinate in addition to the time, one will
encounter inconsistencies when the field equations
(\ref{5d-vacuum1}) and (\ref{5d-vacuum2}) are solved. For example,
these conditions have been used in
Ref.~\cite{aguilar-romero-barros-2007} when the $a=b=c$. Indeed in
this case, for $\varphi(t, l)=\phi(t)\chi(l)$ one has, from
Eq.~(\ref{bulk-eq7}),
\begin{equation}\label{phiequation}
\ddot{\phi}+3H(t)\dot{\phi}+|\lambda|\phi=0
\end{equation}
and
\begin{equation}\label{khi.eq}
\frac{d^2\chi}{dl^2}+|\lambda|\chi(l)=0,
\end{equation}
where, without loss of generality, we have chosen the constant $h$
as $h=1$. The solutions of the above equations are
\begin{equation}\label{solution1timeRomero}
\phi(t)=\phi_0e^{\left(-\frac{3}{2}H_o\pm\frac{1}{2}\sqrt{9H_o^2-4|\lambda|}\,\right)\,t}
\end{equation}
and
\begin{equation}\label{solution2lRomero}
\chi(l)=\chi_0e^{\pm i\sqrt{|\lambda|}\,l}\,,
\end{equation}
where the $H_o$ is a constant Hubble parameter, the $\lambda$ is a
negative separation constant, the $\phi_0$ and $\chi_0$ are
integration constants. However, from Eq.~(\ref{bulk-eq6}) one has
\begin{equation}\label{04component1}
G_{04}=\omega\frac{\dot{\varphi}\varphi'}{\varphi^{2}}+\frac{(\dot{\varphi})'}{\varphi}=0\,,
\end{equation}
where the last step comes from the fact that the $G_{04}$
component vanishes for the assumed metric. Substituting the
separation form of $\varphi$ with solutions
(\ref{solution1timeRomero}) and (\ref{solution2lRomero}) in
Eq.~(\ref{04component1}) yields
\begin{equation}\label{04component3}
(\omega+1)\left(-\frac{3}{2}H_{o}\pm\frac{1}{2}\sqrt{9H_{o}^2-4|\lambda|}\right)\left(\pm
i\sqrt{|\lambda|}\right)=0.
\end{equation}
The relation (\ref{04component3}) for $\omega\neq-1$ gives
$|\lambda|=0$. Therefore, $\chi(l)$ will be a constant, that is,
$\varphi$ does~not depend on the fifth coordinate, and thus
Eqs.~(\ref{bd-energy-momentum}) and (\ref{gen.potential.eq.})
indicate that $T_{\alpha\beta}^{^{(\rm EMT)}}$ and the induced
scalar potential are null contrary to the results obtained in the
literature~\cite{aguilar-romero-barros-2007}. The exception case
is $\omega=-1$ which is precisely the value predicted when the BD
theory is derived as the low energy limit of some string
theories~\cite{Faraoni.book,BD04}.

Also note that, there is a unique solution for when all the metric
components (including the extra component, contrary to the above
condition) depend only on the time while, the scalar field is a
function of the time and the extra coordinate~\cite{Ponce2-10}.

Based on the usual spatial homogeneity, let us now solve the field
equations (\ref{bulk-eq1})--(\ref{bulk-eq7}) using separation of
variables as
\begin{equation}\label{sep.eq}
\varphi(t,l)=\varphi_0 t^{p_{0}}l^{s_{0}},\hspace{6mm}
a(t,l)\propto t^{p_{1}}l^{s_{1}},\hspace{6mm} b(t,l)\propto
t^{p_{2}}l^{s_{2}},\hspace{6mm} c(t,l)\propto
t^{p_{3}}l^{s_{3}},\hspace{6mm} h(t,l)=h_o
t^{p_{4}}l^{s_{4}},\hspace{6mm}
\end{equation}
where the $h_0$ and $\varphi_0$ are constants, the $p_{_A}$'s and
$s_{_A}$'s are parameters satisfying field equations
(\ref{bulk-eq1})--(\ref{bulk-eq7}). Actually, it turns out to be
easier to work with this {\it ansatz}. First substituting {\it
ansatz} (\ref{sep.eq}) into Eq.~(\ref{bulk-eq7}) gives
\begin{equation}\label{phi.eq}
 p_0\left(\sum^{4}_{A=0}p_{_A}-1\right)h_0^2t^{p_0-2}l^{s_0}+
 s_0\left[1+s_4-\left(\sum^{3}_{\mu=0}s_\mu\right)\right]t^{p_0-2p_4}l^{s_0-2-2s_4}=0.
\end{equation}
This relation is satisfied for the following five conditions
\begin{eqnarray}\label{cases}
 {\rm I:}&p_0&=0,\hspace{25mm} s_0=0,\cr
 {\rm II:}&p_0&=1-\sum^{4}_{i=1}p_i,\hspace{11mm}
   s_0=1+s_4-\sum^{3}_{i=1}s_i,\cr
 {\rm III:}&p_0&=0,\hspace{24mm}s_0=1+s_4-\sum^{3}_{i=1}s_i,\cr
 {\rm IV:}&p_0&=1-\sum^{4}_{i=1}p_i,\hspace{11mm} s_0=0,\cr
 {\rm V:}&p_4&=1,\quad s_4=-1,\hspace{5mm} h_0^2p_0
   \left[\left(\sum^{4}_{A=0}p_{_A}\right)-1\right]+s_0\left(1+s_4-\sum^{3}_{\mu=0}s_\mu\right)=0.
 \end{eqnarray}
Clearly, mixtures of these conditions may also occur. The solution
Case~I implies $\varphi$ is a constant which corresponds to
general relativity in five dimensions. Such a solution for the
spatially flat FRW metric has been investigated in
Ref.~\cite{Ponce2-10}. In this work, we purpose to investigate the
Case~II which leads to a generalized Kasner space--time in five
dimensions. Actually, in what follows we solve its corresponding
equations and, in the next section by applying the formalism
presented in the previous section, we investigate the properties
of such a solution on each $4D$ hypersurface.

Now, substituting {\it ansatz} (\ref{sep.eq}) into the rest of the
field equations, namely Eqs.~(\ref{bulk-eq1})--(\ref{bulk-eq6}),
yields
\begin{eqnarray}\label{sep-rel2}
h_0^2\left[p_{1}p_{2}+p_{1}p_{3}+p_{1}p_{4}+p_{2}p_{3}+p_{2}p_{4}+p_{3}p_{4}
-\left(\frac{\omega+2}{2}\right)p_{0}^{2}+p_{0}\right]t^{2p_{4}-2}\cr
=\left[s_{1}^{2}+s_{2}^{2}+s_{3}^{2}+\frac{\omega}{2}s_{0}^{2}-(s_{1}+s_{2}+s_{3})+s_{1}s_{2}
+s_{1}s_{3}+s_{2}s_{3}-s_{4}(s_{1}+s_{2}+s_{3})\right]l^{(-2s_{4}-2)}\,,
\end{eqnarray}
\begin{eqnarray}\label{sep-rel3}
h_0^2\left[p_{2}^{2}+p_{3}^{2}+p_{4}^{2}+\frac{\omega}{2}p_{0}^{2}-(p_{2}+p_{3}+p_{4})-p_{0}p_{1}+
p_{2}p_{3}+p_{2}p_{4}+p_{3}p_{4}\right]t^{2p_{4}-2}\cr
=\left[s_{2}^{2}+s_{3}^{2}+\frac{\omega}{2}s_{0}^{2}-s_{0}s_{1}+s_{2}s_{3}-s_{2}s_{4}-s_{3}s_{4}-s_{2}-s_{3}
\right]l^{(-2s_{4}-2)}\,,
\end{eqnarray}
\begin{eqnarray}\label{sep-rel4}
h_0^2\left[p_{1}^{2}+p_{3}^{2}+p_{4}^{2}+\frac{\omega}{2}p_{0}^{2}-(p_{1}+p_{3}+p_{4})-p_{0}p_{2}+
p_{1}p_{3}+p_{1}p_{4}+p_{3}p_{4}\right]t^{2p_{4}-2}\cr
=\left[s_{1}^{2}+s_{3}^{2}+\frac{\omega}{2}s_{0}^{2}-s_{0}s_{2}+s_{1}s_{3}-s_{1}s_{4}-s_{3}s_{4}-s_{1}-s_{3}
\right]l^{(-2s_{4}-2)}\,,
\end{eqnarray}
\begin{eqnarray}\label{sep-rel5}
h_0^2\left[p_{1}^{2}+p_{2}^{2}+p_{4}^{2}+\frac{\omega}{2}p_{0}^{2}-(p_{1}+p_{2}+p_{4})-p_{0}p_{3}+
p_{1}p_{2}+p_{1}p_{4}+p_{2}p_{4}\right]t^{2p_{4}-2}\cr
=\left[s_{1}^{2}+s_{2}^{2}+\frac{\omega}{2}s_{0}^{2}-s_{0}s_{3}+s_{1}s_{2}-s_{1}s_{4}-s_{2}s_{4}-s_{1}-s_{2}
\right]l^{(-2s_{4}-2)}\,,
\end{eqnarray}
\begin{eqnarray}\label{sep-rel6}
h_0^2\left[p_{1}^{2}+p_{2}^{2}+p_{3}^{2}+\frac{\omega}{2}p_{0}^{2}-(p_{1}+p_{2}+p_{3})-p_{0}p_{4}+
p_{1}p_{2}+p_{1}p_{3}+p_{2}p_{3}\right]t^{2p_{4}-2}\cr
=\left[-\left(\frac{\omega+2}{2}\right)s_{0}^{2}+s_{0}+s_{0}s_{4}+s_{1}s_{2}+s_{1}s_{3}+s_{2}s_{3}
\right]l^{(-2s_{4}-2)}
\end{eqnarray}
and
\begin{equation}\label{sep-rel7}
(\omega+1)p_{0}s_{0}+p_{1}s_{1}+p_{2}s_{2}+p_{3}s_{3}=p_{4}(s_{0}+s_{1}+s_{2}+s_{3}).
\end{equation}
For later on convenience, we will set $h_o=1$ in {\it ansatz}
(\ref{sep.eq}). Also, in order to compare the solutions with the
corresponding ones in the IMT~\cite{H01,PW08}, one must set the BD
scalar field to be a constant. By assuming $p_4\neq 1$ and
$s_4\neq -1$, one can match the coefficients of the left and right
hand sides of relations (\ref{sep-rel2})--(\ref{sep-rel6}). Hence,
after a little manipulation and considering relation
(\ref{sep-rel7}), one finds the following constrains among the
generalized Kasner parameters\footnote{The relations in the second
line of (\ref{cases}), i.e. the Case~II, have been rewritten as
relations (\ref{kas-rel1}) and (\ref{kas-rel3}), respectively.}
\begin{equation}\label{kas-rel1}
\sum^{4}_{A=0}p_{_A}=1\,,
\end{equation}
\begin{equation}\label{kas-rel2}
(\omega+1)p_{0}^{2}+\sum^{4}_{m=1}p_m^{2}=1\,,
\end{equation}
\begin{equation}\label{kas-rel3}
\sum^{3}_{\mu=0}s_{\mu}=1+s_{4}\,,
\end{equation}
\begin{equation}\label{kas-rel4}
(\omega+1)s_{0}^{2}+\sum^{3}_{i=1}s_i^{2}=(1+s_{4})^{2}\,,
\end{equation}
\begin{equation}\label{kas-rel5}
(\omega+1)p_{0}s_{0}+\sum^{3}_{i=1}p_is_i=p_{4}(1+s_{4})\,,
\end{equation}
\begin{equation}\label{kas-pp}
\sum^{4}_{{m, n=1}\atop {m <
n}}p_mp_n=\frac{(\omega+2)}{2}p_{0}^{2}-p_{0}
\end{equation}
and
\begin{equation}\label{kas-ss}
\sum^{3}_{{i, j=1}\atop {i <
j}}s_{i}s_{j}=\frac{(\omega+2)}{2}s_{0}^{2}-s_{0}(1+s_{4}).
\end{equation}
Constrains (\ref{kas-rel1})--(\ref{kas-rel5}) are the generalized
Kasner relations in $5D$ BD theory, though there are only five
independent relations among the Kasner parameters, and relations
(\ref{kas-pp}) and (\ref{kas-ss}) can actually be achieved from
the others.

As it will be shown the case $p_0=0$ has particular properties, and hence, we
will highlight this case in all the following parts of the work.
\begin{itemize}
\item
 In the special case $p_0=0=s_0$,
the BD scalar field is a constant regardless of any specific value
of the BD coupling parameter. Thus,
constrains~(\ref{kas-rel1})--(\ref{kas-rel5}) lead to the
corresponding relations in $5D$ general relativity,
Refs.~\cite{H01,PW08}, which in turn, give the $5D$ analogue of
the well--known Kasner solution.
\item
 When the scale factors and the scalar field only
depend on the time, for the particular case of spatial isotropy,
by assuming $p_1=p_2=p_3\equiv p$, from (\ref{bulk-metric}),
(\ref{sep.eq}), (\ref{kas-rel1}) and (\ref{kas-pp}), one gets
\begin{eqnarray}\label{spec.bulk-metric}
dS^{2}=-dt^{2}+A^2t^{2p}(dx^{2}+dy^{2}+dz^{2})+B^2t^{2p_4}dl^{2},\qquad\qquad{}\\\nonumber
\varphi=\varphi_0t^{1-3p-p_4} \hspace{15mm} {\rm and}
\hspace{15mm} \omega=\frac{2(3p+p_4)(1-p_4)-12p^2}{(1-3p-p_4)^2},
\end{eqnarray}
where the $A$ and $B$ are constants. These results are the same as
the corresponding ones investigated for the spatially flat FRW
solutions in $5D$ BD theory~\cite{Ponce2-10}.

In the limit $p_4\longrightarrow (1-3p)$, one obtains again a
constant $\varphi$ and, in general, an infinite $\omega$. In this
limit\rlap,\footnote{We disregard the static case $p=0$.}\
 only for $p=1/2$, i.e. $p_4=-1/2$, the field equations are
satisfied, namely one has
\begin{eqnarray}\label{spec1.bulk-metric}
dS^{2}=-dt^{2}+A^2t(dx^{2}+dy^{2}+dz^{2})+B^2t^{-1}dl^{2}.
\end{eqnarray}
Solution (\ref{spec1.bulk-metric}) is a unique solution in $5D$
Ricci--flat space--time for metric (\ref{bulk-metric}) in the
particular case where there is~not isotropy in all dimensions.
\end{itemize}

In the next section, by employing the formalism obtained in the
preceding section for the field equations, we derive the induced
scalar potential, the energy density and pressure of the
induced--matter, and hence, discuss the properties of the model.

\section{Reduced BD Cosmology in $4D$ Space--Time}
\indent

For metric (\ref{bulk-metric}), the induced metric on the
hypersurface $\Sigma_0$ is just a part of it, and the
non--vanishing components of induced EMT
(\ref{bd-energy-momentum}) are
\begin{equation}\label{t-00}
\frac{8\pi}{\varphi}T^{0({\rm EMT})}_{\,\,\,0}=
-\frac{\ddot{h}}{h}-\frac{1}{h^2}\left[\frac{a'b'}{ab}+\frac{a'c'}{ac}
+\frac{b'c'}{bc}-\frac{\omega}{2}\left(\frac{\varphi'}{\varphi}\right)^2
+\left(\frac{a'}{a}+\frac{b'}{b}+\frac{c'}{c}
\right)\frac{\varphi'}{\varphi}\right]+\frac{V(\varphi)}{2\varphi}
\end{equation}
and
\begin{equation}\label{t-ii}
\frac{8\pi}{\varphi}T^{1({\rm EMT})}_{\,\,\,1}=
-\frac{\dot{a}\dot{h}}{ah}+\frac{1}{h^2}\left[-\frac{a'h'}{ah}+\frac{a''}{a}
-\frac{b'c'}{bc}+\frac{\omega}{2}\left(\frac{\varphi'}{\varphi}\right)^2
-\left(\frac{b'}{b}+\frac{c'}{c}
\right)\frac{\varphi'}{\varphi}\right]+\frac{V(\varphi)}{2\varphi},
\end{equation}
where the induced potential $V(\varphi)$ should be determined from
(\ref{gen.potential.eq.}). Also, by interchanging
$a\longleftrightarrow b$ and $a\longleftrightarrow c$ in relation
(\ref{t-ii}), one obtains $(8\pi/\varphi)T^{2({\rm
EMT})}_{\,\,\,2}$ and $(8\pi/\varphi)T^{3({\rm EMT})}_{\,\,\,3}$,
respectively. As, in general, the three components $T^{i({\rm
EMT})}_{\,\,\,i}$ (where $i=1,2,3$ with no sum) are~not equal, one
cannot consider the induced--matter as a perfect fluid.

The induced scalar potential is obtained by substituting the power--law relations (\ref{sep.eq})
into definition~(\ref{gen.potential.eq.}) as
\begin{equation}\label{pot.gen1}
\frac{dV}{d\varphi}{\Biggr
|}_{_{\Sigma_{o}}}\!\!\!\!\!=2(\omega+1)p_0p_4t^{-2}+\zeta\,
h_0^{-2}t^{-2p_4}l_0^{-2s_4-2},
\end{equation}
where
\begin{equation}\label{K}
\zeta\equiv -\omega s_0\Big[3s_0+2(s_1+s_2+s_3-s_4-1)\Big]
+2\Big[s_0(s_4-s_0+1)+s_1s_2+s_1s_3+s_2s_3\Big].
\end{equation}
However, constrains~(\ref{kas-rel1})--(\ref{kas-ss}) make
$\zeta=0$ for the model. Therefore, by integrating
(\ref{pot.gen1}) while using the generalized Kasner relations, one
gets
\begin{equation}\label{pot.spec}
V(\varphi)=\left \{
 \begin{array}{c}
 \left[\frac{2(\omega+1)p_0^2p_4}{p_0-2}
 l_0^{\frac{2s_0}{p_0}}\varphi_0^{\frac{2}{p_0}}\right]
 \varphi^{\frac{(p_0-2)}{p_0}}
 \quad {\rm for}\quad p_0\neq0,2\\\nonumber \\
 \Big[4\varphi_0(\omega+1)p_4l_0^{s_0}\Big]
 {\rm ln} \varphi \qquad\quad {\rm for}\quad p_0=2\\
 \end{array}\right.
\end{equation}
on the hypersurface, where we have set the constants of
integration equal to zero. Note that, when $\omega=-1$ and or
$p_4=0$, the effective potential vanishes.
\begin{itemize}
\item
 For the special case $p_0=0$, the $\varphi\frac{dV}{d\varphi}{\Big
|}_{_{\Sigma_{o}}}$ identically vanishes by (\ref{pot.gen1}).
Therefore, without loss of generality, one can set $V=0$ in this
case.
\end{itemize}

In the following subsections we probe the properties of the
induced EMT.

\subsection{Conservation of Induced Energy--Momentum Tensor}
\indent

As mentioned, in the MBDT approach, the induced EMT obeys the
conservation law as in the Einstein theory. Let us check it for
the anisotropic induced EMT.

The conservation law for the induced EMT in the case of Bianchi
type~I line element reduces to
\begin{equation}\label{Cons.eq}
\dot{\rho}+\sum^{3}_{i=1}(\rho+P_i)H_i=0,
\end{equation}
where $\rho\equiv -T^{0({\rm EMT})}_{\,\,\,0}$ and $P_i\equiv
T^{i({\rm EMT})}_{\,\,\,i}$ (no sum on $i$) are the energy density
and the directional pressure of the induced EMT, respectively.

Now, from relations~(\ref{sep.eq}), (\ref{kas-rel3}),
(\ref{kas-ss}), (\ref{t-00}), (\ref{t-ii}) and (\ref{pot.spec})
one obtains
\begin{equation}\label{ro}
\rho=\left \{
 \begin{array}{c}
 \frac{p_4\varphi_0l_0^{s_0}}{8\pi(p_0-2)}
 \Big[(p_0-2)(p_4-1)-(\omega+1)p_0^2\Big]t^{p_0-2} \quad {\rm for}\quad p_0\neq0,2\\\nonumber
 \\
 \frac{p_4\varphi_0l_0^{s_0}}{8\pi}
 \Big[(p_4-1)-2(\omega+1){\rm ln}\left(\varphi_0l_0^{s_0}t^2\right)\Big] \qquad {\rm for}\quad  p_0=2\\
 \end{array}\right.
\end{equation}
and
\begin{equation}\label{P.i}
P_i=\left \{
 \begin{array}{c}
 -\frac{p_4\varphi_0l_0^{s_0}}{8\pi(p_0-2)}
 \Big[(p_0-2)p_i-(\omega+1)p_0^2\Big]t^{p_0-2} \quad {\rm for}\quad  p_0\neq0,2\\\nonumber
 \\
 -\frac{p_4\varphi_0l_0^{s_0}}{8\pi}\Big[p_i-2(\omega+1)
 {\rm ln}\left(\varphi_0l_0^{s_0}t^2\right)\Big] \qquad {\rm for}\quad  p_0=2.\\
 \end{array}\right.
\end{equation}
Obviously, for $i\neq j$, one has $P_i\neq P_j$, that means the
induced EMT is not a perfect fluid. By substituting the
corresponding induced energy density and pressure from relations
(\ref{ro}) and (\ref{P.i}) for the power--law and the logarithmic
forms into (\ref{Cons.eq}) and using relations (\ref{kas-rel1})
and (\ref{kas-rel2}), one can explicitly show that the
conservation law is satisfied. Indeed as mentioned, this result is
expected as a straightforward consequent of the formalism
presented in Section~$2$.
\begin{itemize}
\item
 When $p_0=0$, using the
power--law relations~(\ref{sep.eq}), equations (\ref{t-00}) and
(\ref{t-ii}), and owing to the fact that the induced potential
vanishes for this case, one easily obtains
\begin{equation}\label{spec.ro.ei}
\rho=\frac{\varphi_0l_0^{s_0}}{8\pi}\frac{p_4(p_4-1)}{t^2}
\end{equation}\label{ro.spec}
and
\begin{equation}\label{pi-special}
P_i=-\frac{\varphi_0l_0^{s_0}}{8\pi}\frac{p_4p_i}{t^2}.
\end{equation}\label{p.i.spec}
These results indicate that the conservation law is satisfied for
this case as well.
\end{itemize}

\subsection{Properties of The Physical Quantities}
\indent

In the rest of this work, for
simplicity\rlap,\footnote{Nevertheless, we have confined ourselves
to a fixed hypersurface.}\
 we assume that the components of metric (\ref{bulk-metric}) and the
scalar field only depend on the cosmic time, namely {\it ansatz}
(\ref{sep.eq}) reads as
\begin{equation}\label{spec.sep.eq}
\varphi(t)=\kappa t^{p_{0}},\hspace{8mm}
 a(t)\propto t^{p_{1}},\hspace{8mm}
b(t)\propto t^{p_{2}},\hspace{8mm}
c(t)\propto t^{p_{3}},\hspace{8mm}
h(t)=m t^{p_{4}},\hspace{8mm}
\end{equation}
where the $\kappa$ and $m$ are constants. Now, Eqs.~(\ref{t-00})
and (\ref{t-ii}) lead to
\begin{equation}\label{epec.t-00}
\frac{8\pi}{\varphi}T^{0({\rm EMT})}_{\,\,\,0}=
-\frac{\ddot{h}}{h}+\frac{V(\varphi)}{2\varphi}
\end{equation}
and
\begin{equation}\label{epec.t-11}
\frac{8\pi}{\varphi}T^{i({\rm EMT})}_{\,\,\,i}=
-H_iH_4+\frac{V(\varphi)}{2\varphi}\qquad ({\rm no\ sum\ on}\ i).
\end{equation}
Also, from definition (\ref{gen.potential.eq.}) and relations
(\ref{spec.sep.eq}), one has
\begin{equation}\label{epec.pot}
\varphi\frac{dV}{d\varphi}{\Biggr |}_{_{\Sigma_{o}}}\!\!\!\!\!\!=
2(\omega+1)\kappa^{2/p_0}p_0p_4\varphi^{(p_0-2)/p_0}.
\end{equation}
Integrating the above relation gives
\begin{equation}\label{epec.pot2}
V(\varphi)=
2(\omega+1)\kappa^{2/p_0}\frac{p_0^2p_4}{p_0-2}\varphi^{(p_0-2)/p_0}
\end{equation}
on the hypersurface, where we have set the constant of integration
equal to zero, and have assumed\footnote{From now on, we do~not
investigate the logarithmic potential with $p_0=2$, for it leads
to some difficulties when the weak energy condition ({\bf WEC}) is
applied. For example, in order to satisfy the WEC, one can set
$p_4=0$ when $p_0= 2$, which in turn gives no induced EMT on the
hypersurface.}\
  $p_0\neq0,2$. For the case $p_0=0$, one again has $V=0$. Substituting relation
(\ref{epec.pot2}) into relations (\ref{epec.t-00}) and
(\ref{epec.t-11}) gives
\begin{equation}\label{spec.ro20}
\rho =-T^{0({\rm
EMT})}_{\,\,\,0}=\frac{\kappa}{8\pi}\left(\frac{p_4}{p_0-2}\right)
 \Bigg[(p_0-2)(p_4-1)-(\omega+1)p_0^2\Bigg]t^{p_0-2}
\end{equation}
and
\begin{equation}\label{spec.P.i}
P_i =T^{i({\rm
EMT})}_{\,\,\,i}=-\frac{\kappa}{8\pi}\left(\frac{p_4}{p_0-2}\right)
\Big[(p_0-2)p_i-(\omega+1)p_0^2\Big]t^{p_0-2},
\end{equation}
where we have used relations (\ref{kas-rel1}) and (\ref{kas-rel2}).

For later on convenience, one usually defines the spatial volume
$\bar{V}$, the average scale factor $\alpha$ and the mean Hubble
parameter $H$ as~\cite{CHM01}--\cite{SS10}
\begin{equation}\label{def-Hq}
\bar{V}\equiv abc,\hspace{10mm}\alpha\equiv\sqrt[3]{abc}=\sqrt[3]{\bar{V}} \hspace{10mm} {\rm and} \hspace{10mm}
H\equiv\frac{1}{3}\sum^{3}_{i=1}H_{i}=\frac{1}{3}\frac{\dot{\bar{V}}}{\bar{V}}=\frac{\dot{\alpha}}{\alpha}.
\end{equation}
Hence, by relations  (\ref{kas-rel1}) and (\ref{spec.sep.eq}) one
gets
\begin{equation}\label{ave.scale.Def}
\alpha(t)\propto t^{\beta} \hspace{15mm} {\rm with} \hspace{15mm}
\beta\equiv(p_1+p_2+p_3)/3=(1-p_0-p_4)/3,
\end{equation}
where we exclude $\beta=0$, for it gives a static universe. Let us
also define the effective pressure on the hypersurface as
\begin{equation}\label{gen.eff-pressure}
P_{\rm eff}\equiv\frac{1}{3H}\sum^{3}_{i=1}H_iP_i,
\end{equation}
where this definition yields the conservation law for the
isotropic case. Substituting relation (\ref{spec.P.i}) into
definition (\ref{gen.eff-pressure}) gives
\begin{equation}\label{eff-pressure1}
P_{\rm eff}=-\frac{\kappa}{8\pi}\frac{p_4(1+p_4)}{(p_0-2)}
\left[\frac{(p_0-2)(1-p_4)+(\omega+1)p_0^2}{(1-p_0-p_4)}\right]t^{p_0-2},
\end{equation}
where we have used relation (\ref{kas-rel1}). Hence, relations
(\ref{spec.ro20}) and (\ref{eff-pressure1}) yield a barotropic
equation of state on the hypersurface as $P_{\rm eff}=w\rho$ with
\begin{equation}\label{n}
w=\frac{1+p_4}{1-p_0-p_4}.
\end{equation}
Note that, as $p_0\neq 2$ therefore, $w\neq -1$.

Furthermore, in what follows, we present the properties of a few
important physical quantities such as the expansion scalar
$\Theta$, the shear scalar $\sigma^2$ and the deceleration
parameter $q$ defined in terms of the mean Hubble parameter
respectively as
\begin{equation}\label{Theta}
\Theta=3H\,,
\end{equation}
\begin{equation}\label{shear}
\sigma^2=\frac{1}{2}\sigma_{ij}\sigma^{ij}=
\frac{1}{2}\left(\sum^{3}_{i=1}H_i^2-3H^2\right)
\end{equation}
and
\begin{equation}\label{dec.par}
q=\frac{d}{dt}\left(\frac{1}{H}\right)-1=-\frac{\alpha\ddot{\alpha}}{\dot{\alpha}^2}.
\end{equation}
These physical quantities are important in the observational
cosmology.

Substituting {\it ansatz}~(\ref{sep.eq}) into
relations~(\ref{Theta})--(\ref{dec.par}) gives
\begin{equation}\label{theta-2}
\Theta=\frac{1-p_0-p_4}{t},
\end{equation}
\begin{equation}\label{sigma-2}
\sigma^2=\frac{1}{6t^2}\left[2p_4(1-p_0-2p_4)-(3\omega+4)p_0^2+2(1+p_0)\right]
\end{equation}
and
\begin{equation}\label{gen.q}
q=\frac{2+p_0+p_4}{1-p_0-p_4},
\end{equation}
where again we have used relations (\ref{kas-rel1}) and
(\ref{kas-rel2}). Note that, relation (\ref{gen.q}) indicates that
$q\neq -1$, i.e. the mean Hubble parameter cannot be constant in
this model.

On the other hand, the special law of variation of the Hubble
parameter introduced by Bermann~\cite{B83} gives models with
constant values of the deceleration parameter. Such a type of
assumption has been applied in the literature, see, e.g.,
Refs.~\cite{BG88}--\cite{PAS10}. This type of variation of the
Hubble parameter is consistent with the observation. By appealing
to this law and hence, by considering the deceleration parameter
to be a constant, i.e. $q_0$, one gets
$\alpha(t)=(C_1t+C_2)^{1/(q_0+1)}$, where $C_1\neq0$ and $C_2$ are
the integration constants. With a proper choice of coordinates and
constants of integration, the average scale factor can be written
as\footnote{For convenience, we will drop the index zero from the
constant $q_0$.}\
 $\alpha(t)\propto t^{1/(q+1)}$
that, comparing with relation (\ref{ave.scale.Def}), means
\begin{equation}\label{beta}
 \beta=\frac{1}{q+1}.
\end{equation}
For an expanding universe, one must set $q+1>0$. Thus, any
positive value of $q$ corresponds to a decelerating universe,
while the negative values of $q$, restricted to $-1<q<0$, yield
accelerating ones which are consistent with the recent observed
data~\cite{FT03}. Now, we can set the most useful characters with
respect to a fixed $q$ (as an observational quantity), the state
parameter $w$ and the BD coupling parameter. Indeed, from
relations (\ref{kas-rel1}), (\ref{ave.scale.Def}), (\ref{n}) and
(\ref{beta}), one gets
\begin{equation}\label{p0-p4}
 p_0=\frac{-1+2q-3w}{q+1} \hspace{15mm} {\rm and} \hspace{15mm} p_4=\frac{-(1+q)+3w}{q+1}.
\end{equation}
Substituting $p_0$ and $p_4$ from (\ref{p0-p4}) into relation
(\ref{spec.ro20}) gives
\begin{equation}\label{spec.ro2}
 \rho=\left(\frac{\kappa}{24\pi}\right)\left[\frac{3w-(q+1)}{(1+w)(q+1)^2}\right]
 \left[18w^2+9w(1-2q)-5(1+2q)+4q^2+(1+3w-2q)^2\omega\right]t^{-\frac{3(1+w)}{q+1}}.
\end{equation}
\begin{itemize}
\item
 In the special case $p_{0}=0$, the scalar field is constant and,
regardless of any specific value of the $\omega$, the BD theory
reduces to general relativity. Also, we recover the generalized
Kasner relations of Ref.~\cite{H01}. In this case, by substituting
(\ref{pi-special}) into (\ref{gen.eff-pressure}) and using
relation (\ref{kas-rel1}), one obtains
\begin{equation}\label{special-inuced2}
P_{\rm eff}
=-\frac{\kappa}{8\pi}\frac{p_{4}(1+p_{4})}{t^{2}}.
\end{equation}
Thus, the effective equation of state holds with
\begin{equation}\label{state-eq1}
w=\frac{1+p_4}{1-p_4},
\end{equation}
where (as mentioned before\footnote{In obtaining the generalized
Kasner relations, we assumed this constrain.})
 $p_4\neq 1$, and again $w\neq -1$ (which, in this special case, is also consistent with $q\neq -1$).

In order to satisfy the WEC, one must have $p_4>1$ or $p_4<0$,
where the latter choice indicates that the extra dimension shrinks
by the time. Note that, when $-2<p_4<1$, the deceleration
parameter is positive, and when ($p_4<-2$ or $p_4>1$), it is
negative. For the radiation dominated universe, by setting $w=1/3$
one obtains $p_4=-1/2$ which by substituting it into relations
(\ref{ave.scale.Def}) and (\ref{spec.ro.ei}) and using relation
(\ref{kas-rel1}), gives
\begin{equation}\label{sspec.matt}
\alpha(t)\propto t^{1/2}, \hspace{15mm} \rho(t)\propto
t^{-2}\hspace{15mm} {\rm for} \hspace{15mm} P_{\rm
eff}=\frac{1}{3}\rho.
\end{equation}
For the matter dominated universe, we set $w=0$ which gives
$p_4=-1$. Then, from (\ref{ave.scale.Def}) and (\ref{spec.ro.ei}),
one has
\begin{equation}\label{sspec.matt2}
\alpha(t)\propto t^{2/3}, \hspace{15mm} \rho(t)\propto
t^{-2}\hspace{15mm} {\rm for} \hspace{15mm} P_{\rm eff}=0.
\end{equation}
In both of the above eras, one has $p_4<0$, thus the WEC is
satisfied and the extra dimension shrinks by the cosmic time. Also
for this case, regardless of any specific value of the $\omega$,
the matter and the radiation dominated epoches evolve the same as
the corresponding ones of the flat FRW space--time in general
relativity, as expected.
\end{itemize}

In the following subsections, we probe the properties of two
interesting cases of the radiation dominated era and $w=1$ (known
as the stiff fluid), for $p_0\neq 0, 2$.

\subsubsection {Radiation--Dominated Universe}
\indent

The main motivation of the IMT in deriving the field equations
with matter sources on any $4D$ hypersurface is to interpret the
classical properties of matter (viewed as a manifestation of the
extra dimensions) rather than solely accepting them as a gift. In
particular, when the components of the metric are independent of
the extra coordinate, one obtains the induced EMT to be a
traceless one~\cite{P01}, in another words, a radiation--like
equation of state. Thus, in this subsection, we purpose to
investigate such an equation of state in the MBDT.

Substituting $w=1/3$ into relations (\ref{p0-p4}) and
(\ref{spec.ro2}) gives
\begin{equation}\label{rad.p0}
p_0=\frac{2(q-1)}{q+1},\hspace{20mm}  p_4=-\frac{q}{q+1}
\end{equation}
and
\begin{equation}\label{spec.ro-rad}
 \rho=3P_{\rm eff}=\frac{\kappa q}
 {8\pi(q+1)^2}\left[2q(\omega+2)-q^2(\omega+1)-\omega\right]t^{-\frac{4}{q+1}}.
\end{equation}
Thus, relations (\ref{theta-2}), (\ref{sigma-2}) and
(\ref{epec.pot2}) give
\begin{equation}\label{V-rad}
\Theta={\frac{3}{(q+1)t}},
\end{equation}
\begin{equation}\label{shear-rad}
 \sigma^2=\frac{-4q^2(\omega+1)+2q(4\omega+5)-(4\omega+6)}{2(q+1)^2t^2}
\end{equation}
and
\begin{equation}\label{epec.pot3}
V(\varphi)=\frac{2(\omega+1)
 \kappa^{\frac{q+1}{q-1}}q(q-1)^2}{(q+1)^2}\varphi^{\frac{2}{1-q}}.
\end{equation}
The $\rho>0$ in relation (\ref{spec.ro-rad}) gives $-3/2<\omega<0$
and positive values of $q$ (which is the case for the radiation
dominated epoch)\rlap.\footnote{Note that, there are other ranges
of $\omega$ and $q$ that satisfy the WEC, but the mentioned range
is the only one in which the induced energy density behaves
smoothly at the whole of the range.}\
 When the deceleration
parameter takes positive values, relation (\ref{rad.p0}) implies
that the extra dimension shrinks as the time increases. The
general behavior of the induced energy density at an arbitrary
fixed time is shown versus $q$ and $\omega$ in Figure~$1$.
\begin{figure}
\begin{center}
\epsfig{figure=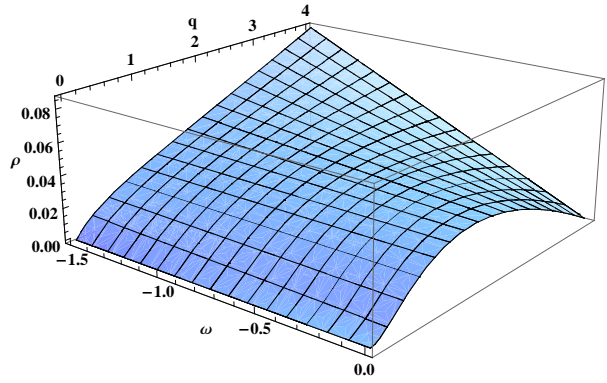,width=8cm}\hspace{5mm}
\end{center}
\caption{\footnotesize The induced energy density at an arbitrary
fixed time as a function of $\omega$ and $q$ in the range of
$-3/2<\omega<0$ and $0<q<4$ for the radiation dominated universe.}
\end{figure}
Also, we have plotted the induced energy density versus time in
Figure~$2$ for an allowed value of $\omega$ and $q$.
\begin{figure}
\begin{center}
\epsfig{figure=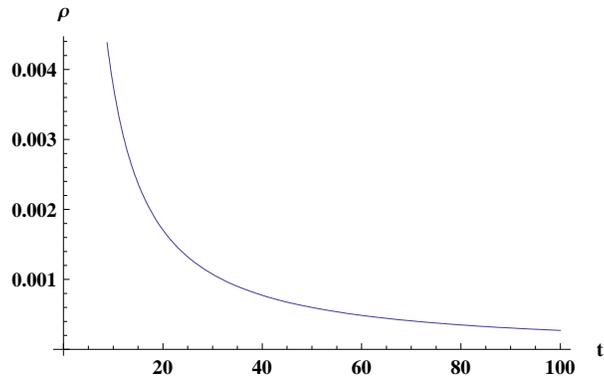,width=8cm}\hspace{5mm}
\end{center}
\caption{\footnotesize
The evolution of the induced energy density in the
 radiation dominated universe with $\omega=-1.1$ and $q=2.5$.}
\end{figure}

By assuming the mentioned range that satisfies the WEC for the
radiation dominated universe, relations
(\ref{spec.ro-rad})--(\ref{epec.pot3}) show that at $t=0$ the
physical quantities $\rho$, $P_{\rm eff}$, $\Theta$, $\sigma^2$
and $V(\varphi)$ take infinite values. However, when
$t\longrightarrow0$ the BD scalar field takes finite values when
$q\geq1$ and diverges when $-1<q<1$. As in the limit
$t\longrightarrow0$, the spatial volume goes to zero and the
expansion scalar tends to infinity, thus the evolution of universe
starts with zero volume and the rate of the expansion is infinite.
These results indicate that the model is in accordance with the
big bang model. As $t$ increases, the expansion scalar decreases
but the spatial volume increases, this shows that the rate of the
expansion slows down by the time. When $t\longrightarrow\infty$,
the spatial volume goes to infinity as well, whereas the other
physical quantities, i.e. the expansion scalar, the scalar
potential, the shear scalar, the induced energy density and the
effective pressure tend to zero. However in this limit, the scalar
field has two limits; it goes to infinity when $q>1$, while it
becomes insignificant when $-1<q<1$. As the ratio
$\sigma^2/\Theta^2$ does~not depend on the time, the model, in
general, does~not approach isotropy for large values of $t$.
Incidentally, the finite value of the integral
\begin{equation}\label{horizon}
 \int_{t_0}^{t}\frac{dt'}{[\bar{V}(t')]^{1/3}}=\frac{q+1}{q}\left[t'^{\frac{q}{q+1}}\right]_{t_0}^{t},
\end{equation}
subject to $q\neq 0$, shows that there is a horizon in this model.

\subsubsection {Stiff Fluid Distribution}
\indent

The high energy density regime of the early universe is usually
described with the stiff cosmological fluid. This kind of fluid,
which is specified by the equation of state $P_{\rm eff}=\rho$,
has been applied for the stellar and cosmological models with
utter dense matter~\cite{Z62}. For the stiff fluid, by
substituting $w=1$ into relations (\ref{p0-p4}) and
(\ref{spec.ro2}), one gains
\begin{equation}\label{stiff.p0}
p_0=\frac{2(q-2)}{q+1},\hspace{20mm}  p_4=\frac{2-q}{q+1}
\end{equation}
and
\begin{equation}\label{spec.ro-stiff}
 P_{\rm eff}=\rho=\frac{\kappa(2-q)}{24\pi(q+1)^2}
 \left[2q^2(\omega+1)-2q(4\omega+7)+(8\omega+11)\right]t^{-\frac{6}{q+1}}.
\end{equation}
Substituting these values into relations~(\ref{theta-2}),
(\ref{sigma-2}) and (\ref{epec.pot2}) gives the physical
quantities as
\begin{equation}\label{V-stiff}
\Theta={\frac{3}{(q+1)t}},
\end{equation}
\begin{equation}\label{shear-stiff}
 \sigma^2=\frac{-2q^2(\omega+1)+(8\omega+11)(q-1)}{(q+1)^2t^2}
\end{equation}
and
\begin{equation}\label{epec.pot3.stiff}
 V(\varphi)=\frac{4\kappa^{\frac{q+1}{q-2}}(\omega+1)(q-2)^3}{3(q+1)^2}\varphi^{\frac{3}{2-q}}.
\end{equation}
For any value of $\omega>-1$ and
$0<q<(4\omega+7-3\sqrt{2\omega+3})/[2(\omega+1)]$, the WEC is
satisfied for the stiff fluid. Figure~$3$ depicts the variation of
the induced energy density at an arbitrary fixed time versus the
BD coupling and the deceleration parameters. Also, in Figure $4$,
the behavior of $\rho$ versus $q$ is plotted at an arbitrary fixed
time and a constant value of the BD coupling parameter. The
evolution of $\rho$, in this case, is similar to the one depicted
in Figure~$2$.
\begin{figure}
\begin{center}
\epsfig{figure=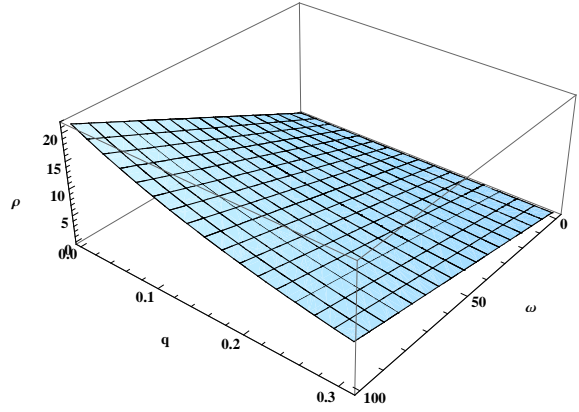,width=8cm}\hspace{5mm}
\end{center}
\caption{\footnotesize The induced energy density at an arbitrary
fixed time versus $\omega$ and $q$ for the stiff fluid, where the
BD coupling parameter is restricted to $-1<\omega<100$.}
\end{figure}
\begin{figure}
\begin{center}
\epsfig{figure=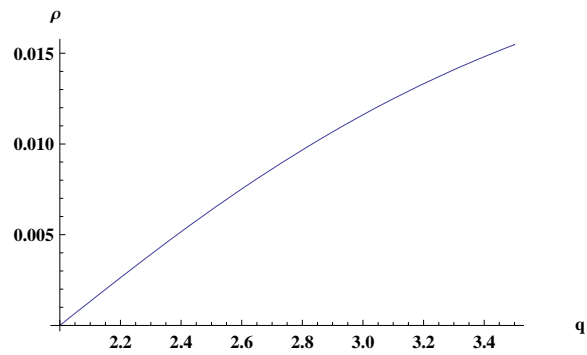,width=8cm}\hspace{5mm}
\end{center}
\caption{\footnotesize The induced energy density at an arbitrary
fixed time versus $q$ with $\omega=0.5$ for the stiff fluid.}
\end{figure}

For this model, one observes that at $t=0$, the spatial volume is
zero while the expansion scalar is infinite, which shows that the
universe starts its evolution with zero volume at the beginning
where the rate of the expansion is infinite. At the initial point
$t=0$, the induced energy density, the effective pressure, the
shear scalar and the induced scalar potential diverge. However,
the BD scalar field has no initial singularity for $q\geq2$ and
diverges for $-1<q<2$. By increasing $t$, the expansion scalar
decreases while the spatial volume increases, which indicates that
the rate of the expansion decreases by the time. When $t$ goes to
infinity the physical quantities $\rho$, $P_{\rm eff}$,
$\sigma^2$, $\varphi$ (for $-1<q<2$), $V(\varphi)$ and $\Theta$
tend to zero, but the spatial volume and the BD scalar field (for
$q>2$) become infinitely large. Thus, in the case of the stiff
fluid, one has an empty universe for large values of the time.
Again, as the ratio $\sigma^2/\Theta^2$ does~not depend on the
time, the model, in general, does~not approach isotropy for large
values of the cosmic time. The integral (\ref{horizon}) also shows
that there is a horizon for the stiff fluid distribution, and
relation (\ref{stiff.p0}) indicates that, for $q>2$, its fifth
dimension shrinks by the time.

\section{Conclusions}
\indent

As the BD theory is a modified version of general relativity, one
can view the MBDT as a modified version of the IMT in which
general relativity is replaced by the BD theory as the fundamental
underlying theory. We purpose to investigate a generalized Bianchi
type~I anisotropic cosmology in $5D$ BD theory. Then, by employing
the obtained formalism, we derive the induced scalar potential,
the energy density and pressure of the induced--matter on any $4D$
hypersurface in the context of the MBDT. Hence, we probe the
properties of the induced quantities for the specified
cosmological model as the quantities fixed by the procedure rather
than introduced {\it priori} by hand. Indeed, the core of the IMT
idea, and hence the MBDT, for inducing $5D$ equations on any $4D$
hypersurface has been motivated to interpret the classical
properties of $4D$ matter rather than solely accepting them.
Hence, in the cosmological applications, what one obtains from
vacuum $5D$ equations on any $4D$ hypersurface is actually the
$5D$ quantities viewed as the ordinary matter.

Meanwhile, we illustrate that if a diagonalized metric, in the
usual spatial dimensional isotropic case and in a gauge with a
constant $g_{44}$, does~not depend on the fifth coordinate while
the scalar field does, it will lead, in general, to
inconsistencies when the field equations are solved. Indeed, the
induced scalar potential and the induced EMT vanish contrary to
the results obtained in the literature. The exception case is
$\omega=-1$ which is precisely the value predicted when the BD
theory is derived as the low energy limit of some string theories.
Also, there is a unique solution for when all the metric
components (including the extra component, contrary to the above
condition) depend only on the cosmic time while the scalar field
is a function of the cosmic time and the extra coordinate.

At first and based on the usual spatial homogeneity, we assume, in
general, that the scale factors of the specified $5D$ metric and
the BD scalar field depend on the time, and also on the extra
coordinate as separated variables in the power--law forms. Thus,
we find five classes of solutions for the generalized Bianchi
type~I geometry in $5D$ space--time through which beside general
investigations, for the sake of compactness, we probe only one
interesting class of them that leads to a generalized Kasner
cosmology in $5D$ BD theory. This class gives a generalized Kasner
relations among the Kasner parameters. One important property of
these relations occurs when the BD scalar field is a constant.
That is, regardless of any specific value of the BD coupling
parameter, the relations reduce to the corresponding relations in
$5D$ general relativity, which in turn, give the $5D$ analogue of
the well--known Kasner solution, as expected. Also, when the scale
factors and the scalar field only depend on the time, in the
particular case of spatial isotropy, the results are the same as
the corresponding ones investigated before for the spatially flat
FRW solutions in $5D$ BD theory.

We then derive the pressure and energy density of the specified
induced--matter on any $4D$ hypersurface. These results indicate
that one cannot consider it as a perfect fluid, however we discuss
the effective $4D$ universe generated by the specified class of
the solutions. The induced scalar potential is found to be either
in the power--law or in the logarithmic form. However, in the
particular case of constant scaler field and even when the scalar
field only depends on the fifth dimension, the induced scalar
potential, without loss of generality, vanishes. Then, the
conservation law (which is supposed to be valid in this MBDT
approach) has been checked explicitly for the derived induced EMT.
Though, as the logarithmic form of the induced EMT has some
difficulties in satisfying the WEC, we do~not investigate it
further.

We proceed our investigations for a few cosmological quantities
where for simplicity (and somehow without loss of generality) we
assume that the metric and the BD scalar field are only functions
of the cosmic time. As usual in the Bianchi type~I theories, we
define a few (mean) parameters and important physical quantities,
however, we have defined the effective pressure in a way to yield
the conservation law in the case of isotropy. The other quantities
under consideration are the spatial volume, the average scale
factor, the mean Hubble parameter, the expansion scalar, the shear
scalar and the deceleration parameter. First we obtain these
quantities in terms of the generalized Kasner parameters. Then, we
find that the induced EMT satisfies the barotropic equation of
state $P_{\rm eff}=w\rho$, where $w$ is a function of the Kasner
parameters. Also, the model indicates that a constant mean Hubble
parameter is~not allowed. Hence, by appealing to the special law
of variation of the Hubble parameter, we assume that the
deceleration parameter to be constant. And thus, we set the
evolution of all the quantities with respect to the fixed
deceleration parameter $q$, the state parameter $w$ and the BD
coupling parameter. In general, the average scale factor indicates
that any positive value of $q$ corresponds to a decelerating
expanding universe, while the negative values of $q$, restricted
to $-1<q<0$, yield accelerating expanding ones.

In the special case of a constant scalar field, regardless of any
specific value of the BD coupling parameter, the BD theory reduces
to general relativity, and we recover the generalized Kasner
relations achieved before in the literature. The model, by
satisfying the WEC, allows a shrinking extra dimension for a
decelerating expanding universe which includes the radiation
dominated and the matter dominated epoches that evolve the same as
the corresponding ones of the flat FRW space--time in general
relativity.

We then probe the quantities, in general case, versus $q$, $t$ and
$\omega$ for the stiff fluid and the radiation dominated universe.
The results show that in the both fluids, one has an expanding
universe commenced with a big bang, and there is a horizon for
each of them. The rate of expansion slows down by the time. Also,
by applying the WEC, the allowed (or the well--behaved) ranges of
the deceleration and the BD coupling parameters have been obtained
for each of the fluids. The behavior of the quantities, in the
very early universe and the very large time, has been discussed.
In particular, the general behavior of the induced energy density,
at an arbitrary fixed time and also its evolution, has been
depicted in a few figures. The models give empty universes when
the cosmic time goes to infinity. However, as the ratio
$\sigma^2/\Theta^2$ does~not depend on the cosmic time, the
models, in general, do~not approach isotropy for large values of
the cosmic time.

As presented, the modified $4D$ BD theory with a non--vanishing
scalar potential and the induced EMT is derived from the $5D$ BD
equations with vacuum configuration, contrary to the standard $4D$
BD theory where the scalar potential introduced by hand. Note
that, though for the Bianchi type~I model interpreted by the
standard approach, introduction of a scalar potential by hand may
provide more freedom in obtaining interesting solutions, but we
have investigated the response of the MBDT approach for the
specified cosmological model.

\section{Acknowledgment}
\indent

We thank the Research Office of the Shahid Beheshti University for
financial support.

\end{document}